# The Evolving Definition of Sepsis

Todd Gary,[1,2] Damian Mingle,[2] and Ashwini Yenamandra[3,4]

[1]Office of Research, Middle Tennessee State University, Murfreesboro, TN; [2]WPC Healthcare, Franklin, TN; [3]Vanderbilt University Medical Center, Nashville, TN; [4]College of Computing and Technology, Lipscomb University, Nashville, TN.

**Abstract**
Sepsis affects millions of people worldwide each year. It occurs when a normal human immune response to a bacterial, viral or fungal infection becomes dysfunctional and triggers widespread inflammation that results in severe tissue damage that leads to organ failure, shock, and death. Sepsis, requires immediate treatment and has a high readmission rate for survivors. It is also one of the most expensive conditions to treat. In 2013, there were more than 1.6 million cases of sepsis in the United States with a financial cost of more than $23 billion. Sepsis was first described in antiquity, and given its current name, by the ancient Greek physician Hippocrates. Despite its long medical history, severity, and financial burden, the causes of sepsis are not well understood, and there is no standard approach to diagnosis and treatment. The definition of sepsis, the characterization of its clinical stages, and sepsis monitoring tools have changed three times in the past 25 years, most recently in March 2016. The universal adoption of this latest definition, sepsis-3, and a screening tool, qSOFA, are currently under debate in the medical community. A means to rapidly identify and treat sepsis could reduce the five million deaths due to sepsis each year worldwide. This paper reviews the evolution of the definition of sepsis and the controversy surrounding the sepsis-3 definition and the sepsis screening tool, qSOFA.

**Abbreviations:**
APACHE: Acute Physiology and Chronic Health Evaluation
EWS: Early Warning Score
HUCP: The Healthcare Cost and Utilization Project
ICU: Intensive Care Unit
MEDS: Mortality in Emergency Department Sepsis
MEWS: Modified Early Warning System
MIMIC: Medical Information Mart for Intensive Care
PIRO: Predisposition, Infection, Response to the infectious challenge, and Organ dysfunction
qSOFA: quick Sepsis-related Organ Failure Assessment
SIRS: Systemic Inflammatory Response Syndrome
SOFA: the Sepsis-related Organ Failure Assessment

# I. Introduction

Sepsis is one of the oldest, costliest, and most devastating medical conditions in the world. Sepsis was first described in ancient Egypt almost 5,000 years ago and given its current name by the Greek physician Hippocrates about 2,400 years ago (Botero and Pérez, 2012). In 2013, the annual financial burden of sepsis in the United States was more than $23 billion. Despite its long history and its current impact on society, sepsis as a medical condition is not well understood. The definition is under current revision (Shankar-Hari et al., 2016; Simpson, 2016; Cortés-Puch and Hartog, 2016; Abraham, 2016); there is no gold standard diagnostic test for sepsis, nor is there a universally adopted treatment (Singer et al., 2016). Along with revising the definition, there is a search for the best screening tool to identify sepsis. A medical task force has utilized the recent availability of large datasets of patients to compare and validate several screening tools for sepsis. These tools were used to explore more than one million patient electronic healthcare records (EHR), which had a known number of patients with sepsis (Seymour, et al., 2016). This use of a data science approach to validating a screening tool for sepsis is a promising area of research. Here we describe the severity of sepsis, its evolving definition, and the screen tool, qSOFA.

### A. The Severity of Sepsis

Sepsis is a significant cause of healthcare costs and mortality rates. Worldwide, an estimated 31.5 million people are treated each year for sepsis of which 5.3 million end up dying from sepsis (Fleischmann, et al., 2016). In the United States, there are about 1.6 million cases of sepsis each year with more than 258,000 deaths, which averages to one person dying of sepsis every two minutes (Sepsis Alliance, 2016). The financial cost of treating sepsis is staggering. The 2015 Healthcare Cost and Utilization Project (HUCP) Statistical Brief places sepsis as the most expensive condition treated in US hospitals at $23.7 billion (Torio and Andrews, 2015). Sepsis accounts for 40 percent of all ICU costs in the United States, and the average cost for treatment of ICU patients with sepsis is six times greater than that for ICU patients that do not have sepsis. Additionally, patients discharged after a serious bout of sepsis have a 62 percent readmission rate (Sutton & Friedman, 2013; Sepsis Alliance, 2014). This information on the severity of sepsis is summarized in figure 1.

**Figure 1:** Notable facts about sepsis.

| Notable | Details |
|---|---|
| High cost to treat | Most expensive condition treated in the United States: $23.7B |
| High ICU costs | 40% of all ICU costs |
| High mortality rate | 60% for septic shock |
| High readmission rate | 62% |
| Rapid decline | Move to septic shock within 36 hours of identification |
| Challenging | No standard diagnostic test, often appears with comorbid conditions |

Research has shown early detection of sepsis is essential in saving lives and reducing cost (Shore, et al., 2007; Jones, et al., 2015; Singer et al., 2016). The Sepsis Alliance estimate that 80% of deaths due to sepsis could be prevented with rapid diagnosis and treatment (Sepsis Alliance, 2016).

## II. The Evolving Definition and Monitoring of Sepsis

The definition of sepsis has shifted over time. Originally sepsis was thought to be an internal rotting or decaying. This was due to the smell of those afflicted (Funk, et al., 2009). Hippocrates applied the Greek word for decay—*sepsis*, meaning decay or to rot—to this medical condition. The development of medical hygiene and the germ theory in the late 1800s by Louis Pasteur, Joseph Lister, and Robert Kock helped change the viewpoint of sepsis from internal decay to originating from a harmful microorganism. In 1914, Hugo Schottmüller laid the foundation for a modern definition of sepsis when he wrote, "Sepsis is present if a focus has developed from which pathogenic bacteria, constantly or periodically, invade the blood stream in such a way that this causes subjective and objective symptoms" (Rittirsch et al., 2008).

As sepsis became more common in ICUs, the role of the patient's immune response became a greater focal point (Bone et al., 1992). In 1992, a conference was held by the Society of Critical Care Medicine and the American College of Chest Physicians to address the lack of consensus regarding the definition of sepsis and the difficulties this created in studies and treatment. This conference and its outcomes are now referred to as sepsis-1. This was followed by sepsis-2 in 2002 and sepsis-3 in 2016.

Key outcomes from the sepsis-1 conference (Bone et al., 1992):

- The establishment of the term "systemic inflammatory response syndrome" (SIRS) was considered a clinical stage of sepsis.
- Key definitions for sepsis were established (see below).
- No definable presence of bacterial infection was required to diagnose sepsis.
- The terms "severe sepsis" and "septic shock" were introduced to differentiate degrees of severity.

Key definitions established at the sepsis-1 conference:

- *Sepsis* results from a host's systemic inflammatory response syndrome (SIRS) to infection.
- *Severe sepsis* is sepsis associated with organ dysfunction, hypoperfusion, or hypotension.
- *Septic shock* is sepsis-induced hypotension persisting despite adequate fluid resuscitation.

The sepsis-1 conference also established four clinical stages or steps. Each stage had specific clinical indicators (see figure 2) as well their challenges for healthcare providers. The first stage is SIRS (systemic inflammatory response syndrome) and is defined by the presence of at least two of the following four indicators: (1) a body temperature above 38.0° C or below 36.0° C; (2) a heart rate above 90 beats/minute; (3) a respiratory rate above 20 breaths per minute; (4) a white blood cell count higher than 12,000 or lower than 4,000 cells per microliter.

If an infection is confirmed or suspected, then the diagnosis changes from SIRS to sepsis. As sepsis progresses from one stage to another, there is a corresponding increase in the mortality rate with septic shock having the highest. Figure 2 lists the 30-day mortality rates published for SIRS and sepsis (Rangel-Fausto et al., 1995) and those published for severe sepsis and septic shock (Schoenberg, et al., 1998).

Each of the four stages listed in figure 2 have been used at one time or another to define sepsis. For example, in ancient times, sepsis was identified near then end of a person's life when untreated the smell of rotting became apparent. This would be known as septic shock today. Then in 1915 the definition of sepsis focused on the infection in stage 2 until this shifted to stage 1, SIRS, in 1989. The most recent definition, sepsis-3, focuses on the beginning of organ failure in stage 3, severe sepsis.

**Figure 2:** Mortality rates within 30 days of diagnosis.

| | Stage 1:<br>**SIRS** | Stage 2:<br>**Sepsis** | Stage 3:<br>**Severe Sepsis** | Stage 4:<br>**Septic Shock** |
|---|---|---|---|---|
| **Mortality Rate** | 7% | 16% | 40% | > 50% |
| **Clinical Indicator** | Flu-like symptoms | SIRS + infection | Sepsis + signs of organ failure | Persistent severe sepsis |
| **When stage was used to define sepsis** | **1989 - 2015**<br>Sepsis-1 &-2 | **1915 - 1989**<br>Pre-Sepsis-1 | **2016 - present**<br>Sepsis-3 | **400BC – 1915**<br>Hippocrates |
| **Challenges for each stage** | Misidentification: inflammation could be due to a wide variety of other medical conditions. | Time: It could take days to verify source of infection. | Treatment: Limited treatment options when organs begin to fail. | High mortality rate: Often too late for successful treatment. |

**A. Sepsis-2 (2001)**

In 2001, a second sepsis definition consensus task force was sponsored by the Society of Critical Care Medicine, European Society of Intensive Care Medicine, American College of Clinical Pharmacy, American Thoracic Society, and the Surgical Infection Society. The consensus task force was formed to review the clinical progress made in treating sepsis and sepsis-1 definition. The task force recommended:

- keeping the sepsis-1 definition, but recognize it has limitations;
- expanding the list of diagnostic criteria for sepsis; and

- recognizing the separate characteristics or stages of sepsis designated by the acronym PIRO: **p**redisposition, **i**nfection, **r**esponse to the infectious challenge, and **o**rgan dysfunction (Levy et al., 2003).

The expanded list of diagnostic criteria of sepsis included 21 bedside or laboratory tests designed to indicate either inflammation or organ dysfunction.

**B. Sepsis Early Warning Systems**

Since the introduction of sepsis-2, there has been an increase in the use of organ failure or mortality scoring systems also known as early warning score (EWS) systems. These EWS systems are used in most ICUs as a method to triage the severity of critically ill patients. They are bedside evaluations designed to provide a likely mortality rate within 28 to 30 days. Several of these EWS systems were developed in line with the sepsis-1 and sepsis-2 definitions.

The EWS systems used in ICUs and adopted to monitor sepsis include:

- APACHE II (the Acute Physiology and Chronic Health Evaluation II) Score, introduced in 1985,
- MEWS (the Modified Early Warning Score), developed in 1999 as a modification of the Morgan's Early Warning Score (Morgan et al., 1997).
- MEDS (see figure 3 and 4), and
- SOFA (the Sepsis-related Organ Failure Assessment) (Vincent, et al., 1996).

To illustrate a EWS, the scores used to determine a patient's mortality rate using MEDS is shown below. Figure 3 is the checklist used to determine the score and figure 4 shows the predicted mortality rate based on the score.

**Figure 3:** The MEDS EWS checklist.

**MEDS Score Checklist**

| | | | |
|---|---|---|---|
| Terminal illness | ☐ No ☐ Yes | 6 points | |
| Age >65 years | ☐ No ☐ Yes | 3 points | |
| Tachypnea or hypoxia | ☐ No ☐ Yes | 3 points | |
| Shock | ☐ No ☐ Yes | 3 points | |
| Platelets <150x10$^9$/L | ☐ No ☐ Yes | 3 points | |
| Altered mental state | ☐ No ☐ Yes | 2 points | |
| Nursing home resident | ☐ No ☐ Yes | 2 points | |
| Lower respiratory infection | ☐ No ☐ Yes | 2 points | |
| MAXIMUM SCORE if all variables present | | 24 points | **TOTAL =** |

**Figure 4:** MEDS' score and corresponding mortality rate.

| Point Range | MEDS Group | Original 28-day Predicted Mortality Rate |
|---|---|---|
| 0 - 4 | Very low | 1.1% |
| 5 - 7 | Low | 4.4% |
| 8 - 12 | Moderate | 9.3% |
| 13 - 15 | High | 16.1% |
| > 15 | Very high | 39.0% |

These EWS systems has helped identify the progress of sepsis in patients and allowed ICUs to determine where to deploy their resources and staff. The use of EWS is significant in the monitoring of sepsis, especially in the ICUs where up to 40 percent of the cost can be due to treating patients with sepsis.

Along with these early warning systems, sepsis awareness campaigns have been shown to be beneficial. The Society of Critical Care Medicine and the European Society of Intensive Care Medicine provide direction for the Surviving Sepsis Campaign, which, in 2009, helped achieve a 25 percent reduction in sepsis mortality through awareness and education about sepsis, especially among healthcare providers (Dellinger et al., 2008). The Sepsis Alliance, a voluntary health organization founded in 2004, is raising awareness of sepsis among healthcare providers and the general public. This included launching sepsis awareness month, September, helping launch the global sepsis alliance, the heroes of sepsis celebration, and designating September 13 as world sepsis day. The heroes of sepsis celebration focuses on recognizing organizations and individuals that have contributed to advancing sepsis research and education. This event brings together survivors with healthcare providers. Damian Mingle, one of the authors of this paper, was a finalist for the 2016 hero of sepsis.

**C. Sepsis-3 (2016)**

In 2014, the European Society of Intensive Care Medicine (ESICM) and the Society of Critical Care Medicine (SCCM) convened a task force of 19 critical care, infectious disease, surgical, and pulmonary specialists with the aim to update the definitions and clinical criteria identifying the "septic patient."

The sepsis-3 task force recognized sepsis as more complex than infection and inflammation and defined sepsis as a "life-threatening organ dysfunction due to a dysregulated host response to infection." In this new definition (sepsis-3), the host response resulting in organ failure from an infection is stressed while the inflammation stage known as SIRS in sepsis-1 and -2 has been removed. The task force included advances made in understanding the pathology, management, and epidemiology of sepsis as well as an analysis of over one million patients. The results published in the February 2016 issue of JAMA, the Journal of the American Medical Association, reduced the clinical stages in sepsis from the four described in figure 2 to the last two stages (Singer et al., 2016). The previous of clinical stage known as severe sepsis is now used to define sepsis. Sepsis is now defined as "a life-threatening organ dysfunction caused by a dysregulated host response to infection." (Singer et al., 2016). The key features and new definition for sepsis are in figure 5 below.

**Figure 5:** Sepsis-3 key features and definitions.

| Key Feature | Definition |
|---|---|
| **Two clinical stages**<br>1. Severe sepsis<br>2. Septic shock<br><br>Introduces a new diagnostic tool, quickSOFA, or qSOFA | **Sepsis** is a life-threatening organ dysfunction caused by a dysregulated host response to infection.<br>**Septic shock** is a subset of sepsis in which particularly profound circulatory, cellular, and metabolic abnormalities are associated with a greater risk of mortality than with sepsis alone. |

To determine which early warning signal to recommend, the sepsis-3 task force compared a variety of EWS systems used to monitor sepsis. The results showed that qSOFA EWS systems was similar to the accuracy to most EWS, but is easier to use and more accurate in locations outside of or prior to admissions into the ICU. Figure 6 has the qSOFA scoring system. A qSOFA score ≥2 suggests a high risk of poor outcome and an indication that these patients have sepsis and should have their lactate levels tested for evidence of organ dysfunction.

**Figure 6:** qSOFA scoring.

| Criteria | Point Value |
|---|---|
| Altered mental status | +1 |
| Respiratory rate > or = 22 | +1 |
| Systolic blood pressure < or = 100 | +1 |

To validate qSOFA as a screening and monitoring tool for sepsis, scores were applied to a large set of electronic health record data which contained known patients with sepsis. The dataset used contained the electronic health record data of 1.3 million encounters from January 1, 2010, to December 31, 2012, at 12 community and academic hospitals within the University of Pittsburgh Medical Center health system in southwestern Pennsylvania were used for this comparison. The predictive validity of qSOFA and other EWS tool were compared by plotting the area under the receiver operating characteristic curve (AUROC). It was the recent availability of a large set of electronic health record data made this study possible.

**D. Controversy surrounding Sepsis-3 definition**

The definition for sepsis developed by the sepsis-3 task force and qSOFA have not been universally adopted. Several papers and letters to the JAMA editor has raised concerns (Sprung & Reinhart, 2016; Kleinpell, et al, 2016; Abraham, E., 2016). A few of the major concerns about the sepsis-3 definition include:

- A focus on adult patients without including newborn and pediatric patients.
- Research generated from patients in the United States and Europe but not from economically poorer countries.
- A task force without experts from emergency medicine and other important groups.
- The change in the clinical stages of sepsis, making comparisons with prior research about sepsis difficult. The past 25-year body of sepsis research focused on four clinical stages of sepsis and the new definition changes this to two clinical stages.
- Too much reliance on qSOFA to diagnose sepsis.
- A lack of endorsement by all societies including the Latin American Sepsis Institute, American College of Chest Physicians, and American College of Emergency Physicians**.**

## III. Area of Promising Research in Sepsis

While there remains controversy surrounding the definition of sepsis, key advances in research continue. One of the areas of promise is using data science approaches to mine insights and knowledge from large databases of electronic healthcare records. This was done to verify and test qSOFA and other EWS. Several hospitals are now utilizing computational means to monitor patients within the ICU. The availability of large sets of digital healthcare records that can be analyzed with data science approaches is able to provide a new approach to the rapid identification and treatment of sepsis.

MIMIC (**M**edical **I**nformation **M**art for **I**ntensive **C**are) is a large database freely available to researchers. It contains heath data from over 40,000 ICU patients who stayed at Beth Israel Deaconess Medical Center between 2001 and 2012. This database contains lab results, electronic documentation, and bedside monitor trends and waveforms. Several research teams have explored the MIMIC database to better understand sepsis (Calvert et al., 2016; Henry et al., 2016). If sepsis can be rapidly detected and treated, the financial cost and patient suffering will decrease.

## IV. Summary

According to the sepsis-3 definition task force, the process of defining sepsis remains a work in progress: "As is done with software and other coding updates, the task force recommends that the new definition be designated Sepsis-3, with the 1991 and 2001 iterations being recognized as Sepsis-1 and Sepsis-2, respectively, to emphasize the need for future iterations." (Singer et al., 2016).

The evolving definition of sepsis can be seen as four major updates to its original definition with needs for future iterations. The original definition began in antiquity when sepsis was named and defined by the rotting smell of its victims. This was updated in 1917 when the definition focused on infection as the trigger for sepsis. In 1992 the sepsis' definition was updated again and this time the focus was on the patient's systemic inflammation response (sepsis-1). In 2016, sepsis' definition was updated a fourth time to focus on organ dysfunction. The lack of universal adaption of the sepsis-3 definition suggests its definition will evolve again as a future iteration suggested by the sepsis-3 task force.

This updating or evolving definition reveals several important factors contributing to the medical condition known as sepsis. These include (1) the pathogen that triggers sepsis; (2) the patient's immune response to the pathogen; and (3) the patient's medical history and condition at the time of sepsis. It is the interplay between these factors that make each case of sepsis to be somewhat unique to the individual and their current medical condition.

In addition to its evolving definition, sepsis is a challenge for healthcare providers to diagnosis and treat. This challenges include:

- The lack of a gold standard for identification and treatment.
- Symptoms of other disorders mask that of sepsis (see appendix A for list).
- A low percentage of patients have sepsis (~2% of emergency room patients have sepsis).
- The rapidly progression of sepsis to a fatal outcome
- A high mortality rate

Sepsis is often not suspected by heath care providers. The clinical symptoms used in diagnosing sepsis (raised temperature, increased breathing rate, and mental alertness) do not rule out other common disorders. In fact, a number of disorders appear as sepsis (see appendix A). Most of these disorders are seen in a higher percentage of patients entering emergency rooms than sepsis. As a result of these challenges, sepsis is often diagnosed too late to save the patient. Although awareness campaigns and ICU early warning systems have been shown to be beneficial.

The rapid identification and treatment of sepsis is critical. In spite of these challenges, several investigators consider sepsis reversible and preventable (Levy, et al., 2004; Marshall, 2015). If sepsis can be rapidly detected and treated, the financial cost and patient suffering will decrease. More exploration of large electronic patient databases involving patients with sepsis, such as MIMIC, using data science approaches are needed. Data science approaches, such as machine learning, predictive analytics, and data mining, are beginning to be applied to difficult healthcare problems (Holzinger and Jurisica, 2014, Chennamsetty, et al., 2015) The result of which should lead to insights into the best ways to identify, monitor and treat for patients with sepsis.

## Acknowledgments

The authors would like to acknowledge the MTSU and WPC Healthcare's Visiting Scholar in Data Science program, the Sepsis Alliance also provided insights, and Jeffry Porter and Sarah Grotelueschen for edits and suggestions on improving the paper.

## Appendix A:
## List of disorder often diagnosed instead of sepsis*

Cholera
Typhoid fever
Salmonella gastroenteritis
Shigella dysenteriae
Staphylococcal food poisoning
Escherichia coli infections
Colitis
Enteritis and gastroenteritis
Tuberculosis
Bubonic plague
Cutaneous anthrax
Brucellosis
Glanders
Melioidosis
Listeriosis
Leprosy
Pulmonary mycobacteria diseases
Faucial diphtheria
Whooping cough
Streptococcal sore throat
Erysipelas
Meningococcal infection
Tetanus
Erythrasma
Gas gangrene
Streptococcus infection
Syphilis
Gonococcal infections
Leptospirosis icterohemorrhagica
Vincent's angina
Yaws
Primary lesions of pinta
Other spirochetal infection
Tinea barbae and tinea capitis
Dermatomycosis
Candidal
Coccidioidomycosis
Histoplasmosis
Blastomycosis
Other mycoses
Opportunistic mycoses
Bacterial meningitis
Meningitis
Intracranial/intraspinal abscess
Pericarditis
Acute and subacute endocarditis
Phlebitis and thrombophlebitis
Acute sinusitis
Acute pharyngitis
Acute tonsillitis
Acute laryngitis and tracheitis
Acute upper respiratory infections
Pneumococcal pneumonia
Other bacterial pneumonia
Bronchopneumonia
Pneumonia
Bronchiectasis
Empyema
Abscess of lung and mediastinum
Appendicitis
Abscess of anal and rectal regions
Peritonitis
Infections of kidney
Urethritis
Acute salpingitis and oophoritis
Inflammatory diseases of prostate
Inflammatory diseases of uterus
Inflammatory disease of cervix
Inflammatory disease of vagina
Inflammatory disease of vulva
Cellulitis and abscess of finger
Acute lymphadenitis, unspecified
Osteomyelitis periostitis
Other functional bladder disorders
Abscess of liver
Portal pyemia
Acute cholecystitis
Urinary tract infection
Pyogenic arthritis
Bacteremia
Postoperative infection
Bronchitis
Diverticulitis
Perforation of intestine

*This list is adapted from Iwashyna et al., 2014.

## Authors information

- **Todd Gary**, PhD is in the Office of Research at Middle Tennessee State University and is a visiting scholar in data science at WPC Healthcare. Email: Todd.Gary@mtsu.edu

- **Damian Mingle**, MBA, is the Chief Data Scientist at WPC Healthcare and was a finalist for the 2016 Sepsis Alliance Hero of the Year. Email: dmingle@wpchealthcare.com

**Ashwini Yenamandra,** PhD, FACMG, Diplomate of American Board of Medical Genetics, is the director of the Clinical Cytogenetics laboratory at Vanderbilt University Medical Center and is a graduate student in the Data Science master's degree program at Lipscomb University. Email: ayenamandra@mail.lipscomb.edu

## Corresponding Author

Damian Mingle
Chief Data Scientist
WPC Healthcare
1802 Williamson Court,
Brentwood, Tennessee, USA
Email: dmingle@wpchealthcare.com